\documentclass[12pt,preprint]{aastex}

\usepackage{emulateapj5}

\usepackage{graphicx}
\usepackage{dcolumn}
\usepackage{bm}
\usepackage{epsf}

\shorttitle{EFFECTS OF DIFFERENTIAL ROTATION}
\shortauthors{LYFORD, BAUMGARTE \& SHAPIRO}

\begin{document}
\title{Effects of Differential Rotation on the Maximum Mass of 
Neutron Stars}
\author{Nicholas D. Lyford\altaffilmark{1},
Thomas W. Baumgarte \altaffilmark{1,2} and
Stuart L. Shapiro \altaffilmark{2,3}}
%
\affil
{\altaffilmark{1} 
Department of Physics and Astronomy, Bowdoin College,
8800 College Station
Brunswick, ME 04011}
\affil
{\altaffilmark{2} 
Department of Physics, University of Illinois at Urbana-Champaign,
Urbana, IL 61801}
\affil
{\altaffilmark{3} 
Department of Astronomy and NCSA, University of Illinois at
Urbana-Champaign, Urbana, IL 61801}
%
\begin{abstract}
The merger of binary neutron stars is likely to lead to differentially
rotating remnants.  In this paper we numerically construct models of
differentially rotating neutron stars in general relativity and
determine their maximum allowed mass.  We model the stars adopting a
polytropic equation of state and tabulate maximum allowed masses as a
function of differential rotation and stiffness of the equation of
state.  We also provide a crude argument that yields a qualitative
estimate of the effect of stiffness and differential rotation on the
maximum allowed mass.
\end{abstract}

\section{Introduction}\label{Intro}

Neutron stars that are newly formed in the coalescence of binary
neutron stars (or in supernova collapse) are likely to be
differentially rotating (see, e.g., the fully dynamical simulations of
Rasio \& Shapiro 1992,1994, Shibata \& Uryu 2000, Faber, Rasio \& Manor 2001, as 
well as the review of Rasio \&
Shapiro 1999 and references therein).  Differential rotation may play
an important role for the stability of these remnants, since it can be
very effective in increasing their maximum allowed mass.

Assuming that neutron stars in binaries have individual masses close
to 1.4 $M_{\odot}$ (Thorsett \& Chakrabarty 1999), and assuming that
the maximum allowed mass of a non-rotating neutron star is in the
range of 1.8 -- 2.3 $M_{\odot}$ (Akmal, Pandharipande \& Ravenhall
1998), one might conclude that the coalescence of binary neutron stars
leads to immediate collapse to a black hole.  However, both thermal
pressure and rotation can increase the maximum allowed mass.

For coalescence from the innermost stable circular orbit, thermal
pressure is believed to have a small effect.  The maximum mass of {\em
uniformly} rotating stars is limited by the spin rate at which the
fluid at the equator moves on a geodesic; any further speed-up would
lead to mass shedding (the ``Kepler'' limit).  Uniform rotation can
therefore increase the maximum allowed mass by at most about 20 \% for
very stiff equations of state (Cook, Shapiro \& Teukolsky, 1992,
hereafter CST1, and 1994, hereafter CST2), which is not sufficient to
stabilize remnants of binary neutron star merger.  Uniformly rotating
equilibrium configurations with rest masses exceeding the maximum rest
mass of nonrotating stars constructed with the same equation of state
are referred to as ``supramassive'' stars (CST1, CST2).

{\em Differential} rotation, however, can be much more efficient in
increasing the maximum allowed mass.  In differentially rotating
stars, the core may rotate faster than the envelope, so that the core
can be supported by rapid rotation without the equator having to
exceed the Kepler limit.  This effect was demonstrated in Newtonian
gravitation by Ostriker, Bodenheimer \& Lynden-Bell (1966) for white
dwarfs, and in general relativity by Baumgarte, Shapiro \& Shibata
(2000, hereafter BSS) for $n=1$ polytropes.  BSS also showed by way of
example that stars with about 60 \% more mass than the maximum allowed
mass of the corresponding non-rotating star can be dynamically stable
against both radial and nonaxisymmetric modes.  BSS refer to
differentially rotating equilibrium configurations with rest masses
exceeding the maximum rest mass of a uniformly rotating star as
``hypermassive'' stars.

In this paper we extend the findings of BSS for $n=1$ polytropes in
two ways.  We survey different polytropic indices, and study the
effect of both differential rotation and stiffness of the equation of
state on the maximum allowed mass.  We also introduce a very simple
model calculation that illustrates these effects qualitatively, and
that moreover gives surprisingly accurate results for soft equations
of state and moderate degrees of differential rotation.

This paper is organized as follows.  In Section \ref{Analytics} we
present qualitative considerations, leading to a simple estimate of
the maximum mass of differentially rotating neutron stars.  In Section
\ref{Numerics} we construct numerical models of fully relativistic,
differentially rotating neutron stars for different polytropic
indices.  We briefly discuss our findings in Section \ref{Discussion}.
We also include Appendix \ref{appA} with tables of our numerical
results.

\section{Qualitative Considerations}
\label{Analytics}

When magnetic fields and relativistic effects can be neglected,
rotating stars in equilibrium satisfy the Newtonian virial theorem
\begin{equation} \label{Virial}
W + 2 T + 3 \Pi = W \left( 1 - 2 \frac{T}{|W|} \right) + 3 \Pi = 0
\end{equation}
(see, e.g., Shapiro \& Teukolsky 1983).  Here the rotational
kinetic energy $T$ scales as 
\begin{equation}
T \sim \frac{J^2}{MR_e^2},
\end{equation}
the potential energy as
\begin{equation}
W \sim - \frac{M^2}{R_e} \sim - M^{5/3} \rho^{1/3},
\end{equation}
and the internal energy, computed from a volume integral of the
pressure, as
\begin{equation} \label{Pi}
\Pi \sim \rho^{1/n} M.
\end{equation}
where $J$ is the angular momentum, $M$ the mass, $R_e$ the equatorial
radius and $\rho$ the mass density.  We have also assumed a polytropic 
equation of state
\begin{equation} \label{eos1}
P = K \rho^{1 + 1/n},
\end{equation}
where $K$ is a constant and $n$ the polytropic index.
Here and throughout we set the gravitation constant
$G=1$.

Inserting the above expressions into the virial equation
(\ref{Virial}), we find
\begin{equation} \label{vir}
-\beta M^{5/3} \rho^{1/3} \left( 1 - 2 \frac{T}{|W|} \right)
+ \gamma \rho^{1/n} M = 0,
\end{equation}
where $\beta$ and $\gamma$ are the appropriate coefficients.  In
general, $\beta$ depends on the eccentricity of the star, but
restricting our analysis to small values of $T/|W|$, and hence to nearly
spherical stars, we assume that both $\beta$ and $\gamma$ are constant.

For non-rotating stars ($T = 0$) we solve equation (\ref{vir})
to find
\begin{equation}
M = \left( \frac{\gamma}{\beta} \right)^{3/2} \rho^{(3 - n)/(2n)},
\end{equation}
while for rotating stars we obtain
\begin{equation} \label{dM1}
M_{\rm rot} = M \left( 1 - 2 \frac{T}{|W|} \right)^{-3/2}.
\end{equation}
For small values of $T/|W|$ the right hand side can be expanded,
and we then find for the mass increase $\delta M \equiv M_{\rm rot} - M$
\begin{equation} \label{dM2}
\frac{\delta M}{M} = 3 \frac{T}{|W|}
\end{equation}
(cf. Shapiro \& Teukolsky 1983, eqn 7.4.40).  This expression determines the
fractional mass increase as a function of $T/|W|$ for a constant value
of the density $\rho$.  We will use this result to estimate the
increase in the maximum allowed mass of a neutron star, even though
typically rotating stars assume their maximum masses at slightly
different densities than the corresponding non-rotating stars.

To evaluate $T$ and $|W|$ we further simplify the problem by assuming
that the star's density profile is a step function with a constant
density $\rho_c$ (equal to the original central density) inside a spherical 
core of radius $R_c$, and zero outside.  From requiring that the mass of this
model star,
\begin{equation}
M = \frac{4 \pi}{3} \rho_c R_c^3,
\end{equation}
be equal to the original mass
\begin{equation}
M = \frac{4 \pi}{3} \bar \rho R_e^3,
\end{equation}
where $\bar \rho$ is the average density, we find the following
relation between the central condensation and the ratio of the radii
\begin{equation} \label{cc}
\frac{R_e}{R_c} = \left( \frac{\rho_c}{\bar \rho} \right)^{1/3}.
\end{equation}
Further assuming that the core is uniformly rotating with the 
central angular velocity $\Omega_c$, we find
\begin{equation}
T = \frac{1}{2} I \Omega_c^2 = \frac{M R_c^2}{5} \Omega_c^2.
\end{equation}
Inserting this relation together with the potential energy
\begin{equation}
W = - \frac{3}{5} \frac{M^2}{R_c}.
\end{equation}
into (\ref{dM2}) yields
\begin{equation} \label{dM3}
\frac{\delta M}{M} \sim \frac{R_c^3 \Omega_c^2}{M}.
\end{equation}
To find the increase in the maximum allowed mass, it is useful
to assume that the star rotates at the mass-shedding limit
\begin{equation}
\Omega_e^2 = \frac{M}{R_e^3},
\end{equation}
where $\Omega_e$ is the equatorial angular velocity.  This equation
can now be used to eliminate the mass $M$ in equation (\ref{dM3}),
which, together with (\ref{cc}), yields
\begin{equation} \label{dM4}
\frac{\delta M}{M} \sim \frac{\bar \rho}{\rho_c} 
	\left( \frac{\Omega_c}{\Omega_e} \right)^2.
\end{equation}

\begin{table*}[t]
\caption{Maximum Mass increase for uniformly rotating polytropes.}  
\centerline{
\begin{tabular}[h]{ccccccc}
        \hline \hline
$n$\tablenotemark{a} &	
$\bar M_0^{\rm max}$ \tablenotemark{b} &
$\bar R^{\rm max}$ \tablenotemark{c} &
$\bar \rho_{\rm max}$  \tablenotemark{d} &
$\delta M/M|_{\rm CST}$ \tablenotemark{e} & 
$\bar \rho / \rho_c |_{\rm TOV}$ \tablenotemark{f} &
$\bar \rho / \rho_c |_{\rm Newt}$ \tablenotemark{g} \\
\hline
0.5	& 0.151	& 0.395	& 1.29	& 0.224	& 0.375	& 0.545 \\
1.0	& 0.180	& 0.763	& 0.42	& 0.146 & 0.209 & 0.304 \\
1.5	& 0.276	& 1.97 	& 0.072	& 0.099 & 0.115 & 0.167 \\
2.0 	& 0.523	& 6.94	& 5.8$\times 10^{-3}$	& 0.066 & 0.063 & 0.088 \\
2.5	& 1.25  & 41.0	& 1.26$\times 10^{-4}$	& 0.040	& 0.034 & 0.043 \\
2.9	& 3.23  & 620	& 1.54$\times 10^{-7}$	& 0.023	& 0.021 & 0.021 \\
\hline
\end{tabular}}
\tablenotetext{a}{Polytropic index.}
\tablenotetext{b}{Maximum rest mass of non-rotating polytrope (CST2).}
\tablenotetext{c}{Circumferential radius of the non-rotating maximum-mass
	configuration (CST2).}
\tablenotetext{d}{Maximum energy density of the non-rotating maximum mass 
	configuration (CST2).}
\tablenotetext{e}{Fractional rest mass increase (CST2).}
\tablenotetext{f}{Estimate (\ref{dM4}) using relativistic central 
	condensation.}
\tablenotetext{g}{Estimate (\ref{dM4}) using Newtonian central 
	condensation.}
\label{table1}
\end{table*}

Equation (\ref{dM4}) provides a very simple estimate for the increase
of the maximum allowed mass.  It depends only on the central
condensation of the non-rotating star, which is a function of the
stiffness of the equation of state, and the ratio of the angular
velocities at the center and equator, which is a function of the
the degree of differential rotation.  For uniformly rotating stars, the maximum mass
increase is estimated to be simply the inverse of the central
condensation.  In Table \ref{table1} we compare this estimate with the
numerical findings of CST2 and find remarkably good
agreement for soft equations of state.  Table \ref{table1} also
illustrates an ambiguity; in Newtonian gravity, the central
condensation is uniquely determined by the polytropic index, but in
general relativity the central condensation of a star depends on the
central density.  We therefore compute a ``relativistic'' central
condensation $\rho_c/\bar \rho|_{\rm TOV}$ from the central energy
density $\rho_c$ and an average density defined as
\begin{equation} \label{avedens}
\bar \rho = \frac{3M}{4\pi R^3}
\end{equation}
of the non-rotating maximum mass model, where $M$ is the total
mass-energy of the star, and $R$ is the circumferential radius.  We
find that this value yields better agreement with the numerical values
of the maximum mass increase than adopting the Newtonian central
condensation.

The ratio $T/|W|$ provides a useful criterion for the onset of secular
($T/|W| \sim 0.14$) or dynamical ($T/|W| \sim 0.27$) non-axisymmetric
instabilities.  Inserting these limits into equation (\ref{dM1}) shows
that mass increases of secularly stable stars are limited by $\delta
M/M \lesssim 1.63$ while mass increases of dynamically stable stars
are limited by $\delta M/M \lesssim 3.2$.  These values agree quite
well with the respective limits of 1.70 and 3.51 found by Shapiro \&
Teukolsky (1983; equations (7.4.41) and (7.4.42)), who also take
stellar deformations into account.


\section{Numerical Results}
\label{Numerics}

We use a modified version of the numerical code of CST1 and CST2 to
construct models of differentially rotating neutron stars.  The code
is based on similar algorithms developed by Hachisu (1986) and
Komatsu, Eriguchi and Hachisu (1989), and we refer to CST1 for
details.  We adopt a polytropic equation of state
\begin{equation} \label{eos2}
P = K \rho_0^{1+1/n},
\end{equation}
where ${\rho_0}$ is the rest-mass density and where equation (\ref{eos2}) 
reduces to equation (\ref{eos1}) in the Newtonian limit.  We
take the polytropic constant $K$ to be unity without loss of
generality.  Since $K^{n/2}$ has units of length, all solutions scale
according to $\bar M = K^{-n/2} M$, $\bar \rho_0 = K^n \rho_0$, etc.,
where the barred quantities are dimensionless quantities
corresponding to $K = 1$, and the unbarred quantities are physical
quantities (compare CST1).

Constructing differentially rotating neutron star models requires 
choosing a rotating law $F(\Omega) = u^t u_{\phi}$, where
$u^t$ and $u_{\phi}$ are components of the four-velocity $u^{\alpha}$
and $\Omega$ is the angular velocity.  We follow CST1 and assume
the rotation law $F(\Omega) = A^2(\Omega_c - \Omega)$, where the
parameter $A$ has units of length.  Expressing $u^t$ and $u_{\phi}$
in terms $\Omega$ and metric potentials yields equation (42) in CST1,
or, in the Newtonian limit,
\begin{equation} \label{diffrotnewt}
\Omega = \frac{\Omega_c}{1 + \hat A^{-2} \hat r^2 \sin^2 \theta}.
\end{equation}
Here we have rescaled $A$ and $r$ in terms of the equatorial radius
$R_e$: $\hat A \equiv A/R_e$ and $\hat r \equiv r/R_e$.  The parameter
$\hat A$ is a measure of the degree of differential rotation and
determines the length scale over which $\Omega$ changes.  Since
uniform rotation is recovered in the limit $\hat A \rightarrow
\infty$, it is convenient to parametrize sequences by $\hat A^{-1}$.
In the Newtonian limit, the ratio $\Omega_c/\Omega_e$ that appears in
the estimate (\ref{dM4}) is related to $\hat A^{-1}$ by
$\Omega_c/\Omega_e = 1 + \hat A^{-2}$, but for relativistic
configurations this relation holds only approximately.


We adopt this particular rotation law for convenience and for easy
comparison with many other authors who have assumed the same law.  We
also compared with the remnants' angular momentum distribution in
the fully relativistic dynamical merger simulations of Shibata and Uryu (2000) and 
to the post-Newtonian simulations of Faber, Rasio and Manor (2001). We have found that 
their numerical results can be fit
reasonably well by our adopted differential rotation law.


We modify the numerical algorithm of CST1 by fixing the maximum
interior density instead of the central density for each model.  This change
allows us to construct higher mass models in some cases, since the
central density does not always coincide with the maximum density and
hence may not specify a model uniquely.  

For a given a value of $n$ and $\hat A$, we construct a sequence of
models for each value of the maximum density by starting with a
static, spherically symmetric star and then decreasing the ratio of
the polar to equatorial radius, $R_{pe} = R_p/R_e$, in decrements of
0.025.  This sequence ends when we reach mass shedding (for large
values of $\hat A$), or when the code fails to converge (indicating
the termination of equilibrium solutions) or when $R_{pe} = 0$ (beyond
which the star would become a toroid).  For each one of these
sequences the maximum achieved mass is recorded.  We repeat this
procedure for different values of the maximum density, covering about
a decade below the central density of the non-rotating maximum mass
model, which yields the maximum allowed mass for the chosen values of
$n$ and $\hat A$.  Our numerical results are tabulated in Appendix
\ref{appA}.  Our maximum mass increases are lower limits in the sense
that even higher mass models may exist, but that we have not been able
to construct them numerically.



\section{Discussion and Summary}
\label{Discussion}

Our numerical results are tabulated in Tables \ref{0.5} to \ref{2.9}
in Appendix \ref{appA}.  We also compare the increases in the maximum
allowed mass with the estimate (\ref{dM4}), and find surprisingly good
agreement for soft equations of state and moderate degrees of
differential rotation.  

In particular, we find that the fractional maximum rest mass increase
$\delta M/M$ for uniformly rotating stars is well approximated by the
inverse of the central concentration (see also Table \ref{table1}).
For moderate degrees of differential rotation, $\delta M/M$ increases
approximately with the square of the ratio between the central and
equatorial angular velocity $\Omega_c/\Omega_e$, in accord with
equation (\ref{dM4}).

For all equations of state we find that $\delta M/M$ increases with
$\Omega_c/\Omega_e$ only up to a moderate value of
$\Omega_c/\Omega_e$, and starts to decrease again for larger values
(at least with our code and algorithm we do not find monotonically
increasing mass configurations).  For stiff equations of state this
turn-around occurs for smaller values of $\Omega_c/\Omega_e$ than for
soft equations of state.  For moderate degrees of differential
rotation $\Omega_c/\Omega_e \lesssim 2$, a given value of
$\Omega_c/\Omega_e$ will lead to a larger increase in the maximum
allowed mass for a stiffer equation of state, as expected from the
estimate (\ref{dM4}).

We find the largest maximum mass increases for moderately stiff
equations of state.  Some of these configurations exceed the maximum
allowed mass of the corresponding non-rotating star by more than a
factor of two.  These configurations typically have large values of
$T/|W| \gtrsim 0.27$, indicating that such stars may by dynamically
unstable against bar formation (but see Shibata, Karino \& Eriguchi
2002, who found mild bar mode instabilities at very small values of
$T/|W|$ for extreme degress of differential rotation).  They are also
``toriodal'', i.e.~assume their maximum density on a torus around the
center of the star, which may indicate an $m=1$ instability at even
smaller values of $T/|W|$ (Centrella, New, Lowe \& Brown 2001).
However, even restricting attention to those configurations that are
not toriodal and have $T/|W| < 0.27$, we find configurations with
masses larger than the maximum mass of the corresponding non-rotating
star by over 60 \%.  BSS demonstrated that at least some of these
models are dynamically stable. Shibata \& Uryu (2000) demonstrated
that binary mergers may result in similarly stable hypermassive stars,
when the progenitor masses are not too close to the maximum mass.

To summarize, we find that differential rotation is very effective in
increasing the maximum allowed mass, especially for moderately stiff
equations of state.  The effect is probably large enough to stabilize
the remnants of binary neutron star merger, which are likely to be
differentially rotating.  Binary neutron star coalescence may
therefore lead to secularly stable, hypermassive neutron stars.  As
discussed in BSS (see also Shapiro 2000), magnetic braking is likely
to bring such differentially rotating stars into uniform rotation,
which reduces the maximum allowed mass and induces a delayed collapse
to a Kerr black hole.

\acknowledgments

This work was supported in part by NSF Grants PHY-0090310 and
PHY-0205155 and NASA Grant NAG5-10781 at the University of Illinois at
Urbana-Champaign and NSF Grant PHY-0139907 at Bowdoin College.  NDL
gratefully acknowledges support through an Undergraduate Research
Assistantship from the Department of Physics and Astronomy at Bowdoin
College.

\begin{appendix}

\section{Numerical Results for Maximum Masses}
\label{appA}

We list below in Tables \ref{0.5} to \ref{2.9} values for the maximum
rest mass increase for uniformly and differentially rotating
polytropes.  For each polytropic index $n$ we tabulate the
differential rotation parameter $\hat A^{-1}$, the ratio of the
central and equatorial angular velocity $\Omega_c/\Omega_e$ (which
reduces to (\ref{diffrotnewt}) in the Newtonian limit, i.e.~for soft
equations of state), the numerically determined fractional rest mass
increase $(\delta M/M)_{\rm num}$, the ratio of the (relativistic)
rotational kinetic energy and the gravitational binding energy
$T/|W|$, the ratio between polar and equatorial radius $R_p/R_e$, the
maximum density $\bar \rho_{\rm max}$, and the estimate $(\delta
M/M)_{\rm est}$ according to equation (\ref{dM4}).  In these estimates
we used the numerically determined ratios $\Omega_c/\Omega_e$ and the
central condensations according to (\ref{avedens}).  For $n \leq 1.25$
some of the maximum mass configurations are toriodal, i.e.~assume the
maximum density on a toroid about the center.  For these polytropic
indices we also include the ratio $\rho_c/\rho_{\rm max}$.

All models are computed with the code of CST1 and CST2, using 64 zones
both in the radial and angular direction, and truncating the Legendre
polynomial expansion at $\ell = 16$ (see CST1 for details of the
numerical implementation).  The accuracy of individual stellar models
can be tested, for example, by computing a relativistic Virial theorem
(Gourgoulhon \& Bonazzola 1994; Cook, Shapiro \& Teukolsky 1996; see
also Nozawa et al.~1998 for a comparison of several different
computational methods).  In our analysis, however, the error in the
maximum mass and related quantities is dominated by the finite step
size in the sequences over $R_p/R_e$ and $\bar \rho_{\rm max}$, which
result in errors typically in the order of a few percent.  For soft
equations of state the mass as a function of central density is a very
slowly varying function, making it quite difficult to determine the
central density of the maximum mass configuration very accurately.
The error in the $R_p/R_e$ is determined by our stepsize of 0.025.  We
finally note that highly toroidal configurations depend very
sensitively on the input parameters, so that those mass increases
should only be taken as estimates.

\begin{table}[!h]
\caption{$n=0.5$: 
$M_0^{\rm max} = 0.151$; $\bar \rho/\rho_c|_{\rm TOV} = 0.375$.}
\centerline{
\begin{tabular}[!htb]{cccccccc}
        \hline \hline
$\hat{A}^{-1}$ & $\Omega_c/\Omega_e$ & $(\delta M/M)_{\rm num}$ & 
$T/|W|$ & $R_p/R_e$ & $\bar \rho_{\rm max}$ 
& $\rho_{\rm max}/\rho_c$ & $(\delta M/M)_{\rm est}$ \\
        \hline 
0.0 	& 1.00	& 0.22 & 0.15	& 0.55	& 1.02 & 1	& 0.38 \\
0.3 	& 1.51	& 0.41 & 0.23	& 0.425	& 0.73 & 1	& 0.86 \\
0.5 	& 1.93	& 0.62 & 0.31	& 0.2	& 0.30 & 0.77	& 1.40 \\
0.7 	& 2.46	& 0.46 & 0.30	& 0.025	& 0.29 & 0.085	& 2.27 \\
1.0 	& 3.54	& 0.20 & 0.27	& 0.1	& 0.31 & 0.29	& 4.70 \\
\hline
\end{tabular}}
\renewcommand{\baselinestretch}{1}
\label{0.5}
\end{table}

\begin{table}[!h]
\caption{$n=0.75$: 
$M_0^{\rm max} = 0.159$; $\bar \rho/\rho_c|_{\rm TOV} = 0.28$.}
\centerline{
\begin{tabular}[!htb]{cccccccc}
        \hline \hline
$\hat{A}^{-1}$ & $\Omega_c/\Omega_e$ & $(\delta M/M)_{\rm num}$ & 
$T/|W|$ & $R_p/R_e$ & $\bar \rho_{\rm max}$ 
& $\rho_{\rm max}/\rho_c$ & $(\delta M/M)_{\rm est}$ \\
        \hline 
0.0 	& 1.00	& 0.18 	& 0.11	& 0.575	& 0.67 	& 1	& 0.28 \\
0.3 	& 1.35	& 0.27 	& 0.15	& 0.5	& 0.64 	& 1	& 0.52 \\
0.5 	& 1.92	& 0.51 	& 0.22	& 0.4	& 0.42 	& 1	& 1.04 \\
0.7 	& 2.49	& 1.07 	& 0.30	& 0.025	& 0.16 	& 0.021	& 1.75 \\
1.0 	& 3.48	& 0.68 	& 0.27	& 0.025	& 0.16  & 0.019 & 3.42 \\
1.5 	& 6.33	& 0.19 	& 0.16	& 0.425	& 0.27 	& 0.74	& 11.3 \\
\hline
\end{tabular}}
\label{0.75}
\renewcommand{\baselinestretch}{1}
\end{table}

\begin{table}[!h] 
\caption{$n=1$: 
$M_0^{\rm max} = 0.180$; $\bar \rho/\rho_c|_{\rm TOV} = 0.209$.}
\centerline{
\begin{tabular}[!htb]{cccccccc}
        \hline \hline
$\hat{A}^{-1}$ & $\Omega_c/\Omega_e$ & $(\delta M/M)_{\rm num}$ & 
$T/|W|$ & $R_p/R_e$ & $\bar \rho_{\rm max}$ 
& $\rho_{\rm max}/\rho_c$ & $(\delta M/M)_{\rm est}$ \\
        \hline 
0.0 	& 1.00	& 0.15 	& 0.083	& 0.575	& 0.35 	& 1	& 0.21 \\
0.3 	& 1.24	& 0.20 	& 0.10	& 0.55	& 0.33 	& 1	& 0.32 \\
0.5 	& 1.65	& 0.31 	& 0.14	& 0.475	& 0.32 	& 1	& 0.57 \\
0.7 	& 2.33	& 0.61 	& 0.21	& 0.375	& 0.23 	& 1	& 1.14 \\
0.8	& 2.66 	& 1.12	& 0.28	& 0.25	& 0.083	& 0.65	& 1.48 \\
0.85	& 2.78	& 1.40	& 0.29	& 0.025	& 0.068	& 5.0$\times 10^{-3}$ & 1.60 \\
1.0 	& 3.39	& 1.22 	& 0.28	& 0.025	& 0.075 & 4.7$\times 10^{-3}$ & 2.28 \\
1.5 	& 6.33	& 0.31 	& 0.15	& 0.475	& 0.23 	& 0.81	& 8.37 \\
\hline
\end{tabular}}
\label{1.0}
\renewcommand{\baselinestretch}{1}
\end{table}

\begin{table}[!h] 
\caption{$n=1.25$: 
$M_0^{\rm max} = 0.216$; $\bar \rho/\rho_c|_{\rm TOV} = 0.15$.}
\centerline{
\begin{tabular}[!htb]{cccccccc}
        \hline \hline
$\hat{A}^{-1}$ & $\Omega_c/\Omega_e$ & $(\delta M/M)_{\rm num}$ & 
$T/|W|$ & $R_p/R_e$ & $\bar \rho_{\rm max}$ 
& $\rho_{\rm max}/\rho_c$ & $(\delta M/M)_{\rm est}$ \\
        \hline 
0.0 	& 1.00	& 0.12 	& 0.063	& 0.6	& 0.15 	& 1	& 0.15 \\
0.3 	& 1.19	& 0.16 	& 0.075	& 0.575	& 0.15 	& 1	& 0.21 \\
0.5 	& 1.51	& 0.21 	& 0.10	& 0.525	& 0.15 	& 1	& 0.33 \\
0.7 	& 1.98	& 0.32 	& 0.13	& 0.475	& 0.14 	& 1	& 0.58 \\
1.0 	& 3.39	& 1.78 	& 0.28	& 0.025	& 0.037 & 1.1$\times 10^{-3}$ & 1.69 \\
1.5 	& 5.26	& 0.27 	& 0.12	& 0.575	& 0.101	& 0.99	& 4.10 \\
\hline
\end{tabular}}
\label{1.25}
\renewcommand{\baselinestretch}{1}
\end{table}

\begin{table}[!h] 
\caption{$n=1.5$: 
$M_0^{\rm max} = 0.276$; $\bar \rho/\rho_c|_{\rm TOV} = 0.115$.}
\centerline{
\begin{tabular}[!htb]{ccccccc}
        \hline \hline
$\hat{A}^{-1}$ & $\Omega_c/\Omega_e$ & $(\delta M/M)_{\rm num}$ & 
$T/|W|$ & $R_p/R_e$ & $\bar \rho_{\rm max}$ 
& $(\delta M/M)_{\rm est}$ \\
        \hline 
0.0 	& 1.00	& 0.10 & 0.047	& 0.625	& 0.061 & 0.12 \\
0.3 	& 1.15	& 0.12 & 0.055	& 0.6	& 0.060 & 0.15 \\
0.5 	& 1.42	& 0.16 & 0.068	& 0.575	& 0.059 & 0.23 \\
0.7 	& 1.81	& 0.22 & 0.089	& 0.525	& 0.056 & 0.38 \\
1.0 	& 2.65	& 0.40 & 0.137	& 0.45	& 0.047 & 0.81 \\
1.5 	& 4.75	& 0.25 & 0.098	& 0.625	& 0.050 & 2.60 \\
\hline
\end{tabular}}
\renewcommand{\baselinestretch}{1}
\label{1.5}
\end{table}

\begin{table}[!h] 
\caption{$n=2.0$: 
$M_0^{\rm max} = 0.523$; $\bar \rho/\rho_c|_{\rm TOV} = 0.063$.}
\centerline{
\begin{tabular}[!htb]{ccccccc}
        \hline \hline
$\hat{A}^{-1}$ & $\Omega_c/\Omega_e$ & $(\delta M/M)_{\rm num}$ & 
$T/|W|$ & $R_p/R_e$ & $\bar \rho_{\rm max} \times 10^3$ 
& $(\delta M/M)_{\rm est}$ \\
        \hline 
0.0 	& 1.00	& 0.067 & 0.027	& 0.65	& 5.1 	& 0.063 \\
0.3 	& 1.12	& 0.076 & 0.031	& 0.625	& 5.1 	& 0.079 \\
0.5 	& 1.33	& 0.092 & 0.036	& 0.625	& 5.1 	& 0.11 \\
0.7 	& 1.63	& 0.12 	& 0.045	& 0.6	& 4.9 	& 0.17 \\
1.0 	& 2.28	& 0.18 	& 0.064	& 0.55	& 4.4 	& 0.33 \\
1.5 	& 4.03	& 0.15 	& 0.059	& 0.7	& 7.0 	& 1.03 \\
\hline
\end{tabular}}
\renewcommand{\baselinestretch}{1}
\label {2.0}
\end{table}

\begin{table}[!h] 
\caption{$n=2.5$: 
$M_0^{\rm max} = 1.25$; $\bar \rho/\rho_c|_{\rm TOV} = 0.034$.}
\centerline{
\begin{tabular}[!htb]{ccccccc}
        \hline \hline
$\hat{A}^{-1}$ & $\Omega_c/\Omega_e$ & $(\delta M/M)_{\rm num}$ & 
$T/|W|$ & $R_p/R_e$ & $\bar \rho_{\rm max} \times 10^4$ 
& $(\delta M/M)_{\rm est}$ \\
        \hline 
0.0 	& 1.00	& 0.043 & 0.016	& 0.675	& 1.15 	& 0.034 \\
0.3 	& 1.10	& 0.048 & 0.017	& 0.65	& 1.15 	& 0.041 \\
0.5 	& 1.28	& 0.056 & 0.020	& 0.65	& 1.1 	& 0.056 \\
0.7 	& 1.54	& 0.069 & 0.024	& 0.625	& 1.1 	& 0.081 \\
1.0 	& 2.10	& 0.098 & 0.034	& 0.6	& 1.1 	& 0.15 \\
1.5 	& 3.50	& 0.102 & 0.035	& 0.75	& 1.05 	& 0.42 \\
2.0	& 5.44	& 0.053	& 0.019	& 0.875	& 1.1	& 1.01 \\
\hline
\end{tabular}}
\renewcommand{\baselinestretch}{1}
\label{2.5}
\end{table}

\begin{table}[!h]
\caption{$n=2.9$: 
$M_0^{\rm max} = 3.23$; $\bar \rho/\rho_c|_{\rm TOV} = 0.021$.}
\centerline{
\begin{tabular}[!htb]{ccccccc}
        \hline \hline
$\hat{A}^{-1}$ & $\Omega_c/\Omega_e$ & $(\delta M/M)_{\rm num}$ & 
$T/|W|$ & $R_p/R_e$ & $\bar \rho_{\rm max} \times 10^7$ 
& $(\delta M/M)_{\rm est}$ \\
        \hline 
0.0 	& 1.00	& 0.028 & 0.010	& 0.675	& 1.4 	& 0.021 \\
0.3 	& 1.09	& 0.031 & 0.011	& 0.675	& 1.3 	& 0.025 \\
0.5 	& 1.25	& 0.037 & 0.013	& 0.675	& 1.3 	& 0.033 \\
0.7 	& 1.50	& 0.044 & 0.015	& 0.65	& 1.3 	& 0.047 \\
1.0 	& 2.02	& 0.062 & 0.020	& 0.625	& 1.3 	& 0.085 \\
1.5 	& 3.29	& 0.069 & 0.022	& 0.775	& 1.3 	& 0.227 \\
2.0 	& 5.07	& 0.035 & 0.012	& 0.9	& 1.4 	& 0.540 \\
\hline
\end{tabular}}
\renewcommand{\baselinestretch}{1}
\label{2.9}
\end{table}

\end{appendix}

\end{document}